\documentclass[12pt]{iopart}

\usepackage{iopams}

\usepackage[utf8]{inputenc}
\usepackage{graphicx}
\usepackage{listings}
\usepackage{xcolor}
\usepackage{multirow}
\usepackage{braket}

\usepackage{array}
\newcolumntype{P}[1]{>{\raggedright\arraybackslash}p{#1}}

\usepackage{subfigure}

\usepackage{hyperref}
\usepackage{xurl} % allows line breaks in long URLs

\begin{document}

\title[Introducing QC into Statistical Physics]{Introducing Quantum Computing into Statistical Physics: Random Walks and the Ising Model with Qiskit}

\author{Zihan Li$^1$, Dan A.\ Mazilu$^1$, Irina Mazilu$^{1*}$}
\address{$^1$Department of Physics and Engineering, Washington and Lee University, Lexington, VA 24450, USA}
\ead{zli@mail.wlu.edu, mazilud@wlu.edu, mazilui@wlu.edu}

\begin{indented}
\item[] $^*$Corresponding author: zli@mail.wlu.edu
\end{indented}

% \submitto{\EJP}

\begin{abstract}

Quantum computing offers a powerful new perspective on probabilistic and collective behaviors traditionally taught in statistical physics. This paper presents two classroom-ready modules that integrate quantum computing into the undergraduate curriculum using Qiskit: the quantum random walk and the Ising model. Both modules allow students to simulate and contrast classical and quantum systems, deepening their understanding of concepts such as superposition, interference, and statistical distributions. We outline the quantum circuits involved, provide sample code and student activities, and discuss how each example can be used to enhance student engagement with statistical physics. These modules are suitable for integration into courses in statistical mechanics, modern physics, or as part of an introductory unit on quantum computing.

\end{abstract}

\maketitle % title page is now complete

\section{Introduction} % Section titles are automatically converted to all-caps.
% Section numbering is automatic.

Quantum computing is a transformative field that is reshaping the landscape of science and technology. By combining ideas from physics, engineering, computer science, and mathematics, it has the potential to address complex problems, from secure communication \cite{gisin2002,shor1994} to medical breakthroughs \cite{biamonte2017,cao2019}. As quantum computing progresses \cite{preskill2018}, introducing its concepts in undergraduate physics courses offers students an early opportunity to explore the principles of quantum mechanics in a practical context. This approach can deepen their understanding of core physics ideas while also highlighting how quantum phenomena drive new technologies. Integrating these ideas into the undergraduate curriculum, rather than reserving them for  graduate-level courses, provides students with a timely and accessible introduction to contemporary scientific advancements \cite{perron2023}. Quantum walks and spin systems like the Ising model not only illustrate key ideas such as superposition and interference, but also connect to foundational topics like diffusion, phase transitions, and optimization.

In this paper, we present two classroom-ready modules designed to integrate quantum computing into the statistical physics curriculum. These modules focus on the quantum random walk and the Ising model, providing opportunities for students to contrast classical and quantum approaches through simulation and code-based activities. Implemented using Qiskit within Jupyter notebooks\cite{Jupyter}, the activities are accessible to students with basic knowledge of quantum mechanics and Python programming. 

The Jupyter notebooks\cite{Jupyter} paired with these models serve as essential pedagogical
tools for instructors and students alike, and are not merely supplementary appendices. These interactive notebooks allow students to actively explore and reinforce theoretical quantum computing
concepts presented in the paper. By working through these examples, students gain hands-on experience with quantum circuits, deepen their conceptual understanding of probability distributions and collective behavior, and encounter quantum computing tools now emerging in scientific research and industry. Although designed for a statistical physics course, the modules can also be adapted for use in modern physics, quantum mechanics, or introductory quantum computing courses.

The remainder of the paper is organized as follows. Section~\ref{sec:qwalk} explores the classical and quantum random walk, including an analytical derivation and circuit-based simulation. Section~\ref{sec:ising} presents the Ising model in both one and two dimensions, highlighting an optimized quantum implementation. Section~\ref{sec:conclusion} offers conclusions and possible extensions.

\section{Classical and Quantum Random Walks}
\label{sec:qwalk}

Before exploring quantum random walks, students need a working understanding of qubits, superposition, and quantum measurement. In statistical physics, students often encounter simple two-state systems, like a coin toss (heads or tails) or the spin of a particle, which can be either `up' or `down'. In the quantum world, things get more intriguing. A quantum bit, or qubit, doesn't follow this straightforward `either-or' rule. Unlike a classical bit, which is firmly either 0 or 1, a qubit can exist in a superposition of both states at once. It's tempting to imagine a coin twirling through the air, seemingly suspended between heads and tails, but this is only a loose analogy. The spinning coin is always in one state; we just don't know it yet. A qubit, in contrast, truly exists in a blend of both until measured. This ability to hold multiple states simultaneously is what gives qubits an edge in processing complex information, making them uniquely powerful for certain kinds of computations.

In a classical random walk, the walker's direction is determined by a coin flip, yielding binary outcomes: heads or tails. In the quantum version, the coin is replaced by a qubit, which behaves more like a spin state than a coin. Unlike classical coin states, spin states (and hence qubits) can exist in superposition and evolve unitarily under quantum operations. This capacity for superposition and interference is what gives quantum walks their distinctive behavior.

To represent quantum states, we use Dirac notation \cite{griffiths2018}, where we write states like \( |\uparrow\rangle \) for `spin-up' and \( |\downarrow\rangle \) for `spin-down'. Using Dirac notation for qubit states  in a course other than quantum mechanics  can deepen students' understanding of concepts like measurement probability, superposition, and quantum interference, distinctions that make quantum simulations particularly powerful. In quantum computing, we often use \( |0\rangle \) and \( |1\rangle \) to represent these states, with \( |0\rangle \) corresponding to spin-up and \( |1\rangle \) to spin-down. Classically, we would expect a particle to be in one of these states, much like a classical bit that is either 0 or 1.

However, quantum mechanics allows particles, like qubits, to exist in a superposition of these states. The general state of a spin-\(\frac{1}{2}\) particle can be written as:
\begin{equation}
|\psi\rangle = \alpha |\uparrow\rangle + \beta |\downarrow\rangle
\end{equation}
where \( \alpha \) and \( \beta \) are complex numbers related to the probabilities of each state. The probability of finding the particle in the spin-up state is \( |\alpha|^2 \), and in the spin-down state is \( |\beta|^2 \). These probabilities must add up to 1, giving us the normalization condition:
\begin{equation}
|\alpha|^2 + |\beta|^2 = 1
\end{equation}
IBM Quantum Composer \cite{ibm_quantum} is an excellent tool for visualizing a qubit's state and can make abstract concepts like superposition more accessible. Instructors can have students experiment with a single qubit in Quantum Composer, using visualizations on the Bloch sphere to observe how the qubit's state changes.

We introduce the concept of a quantum coin implemented via the Hadamard gate, which transforms the basis state \( |0\rangle \) into an equal superposition:
\begin{equation}
H|0\rangle = \frac{1}{\sqrt{2}}(|0\rangle + |1\rangle),
\end{equation}
making it the quantum analog of a fair coin toss. This operation forms the basis of the quantum random walk, where each step depends on the outcome of a quantum `coin flip'.

A sample circuit in the accompanying Jupyter notebook\cite{Jupyter} uses Qiskit to simulate the quantum coin toss. By executing the circuit multiple times with the Aer simulator, students can collect outcome statistics and compare them to the deterministic behavior of classical random walks. This initial hands-on activity prepares students for understanding how quantum randomness and interference modify diffusion and probability distributions.

Random walks provide a familiar and powerful framework for introducing statistical concepts in physics. They help students understand how seemingly random steps can lead to predictable aggregate behavior, such as Gaussian distributions. In this module, we build on this foundation by contrasting classical and quantum random walks—two models that differ significantly in their underlying rules, yet offer a rich opportunity for comparison.

\begin{figure}
    \centering
    \includegraphics[width=0.8\textwidth]{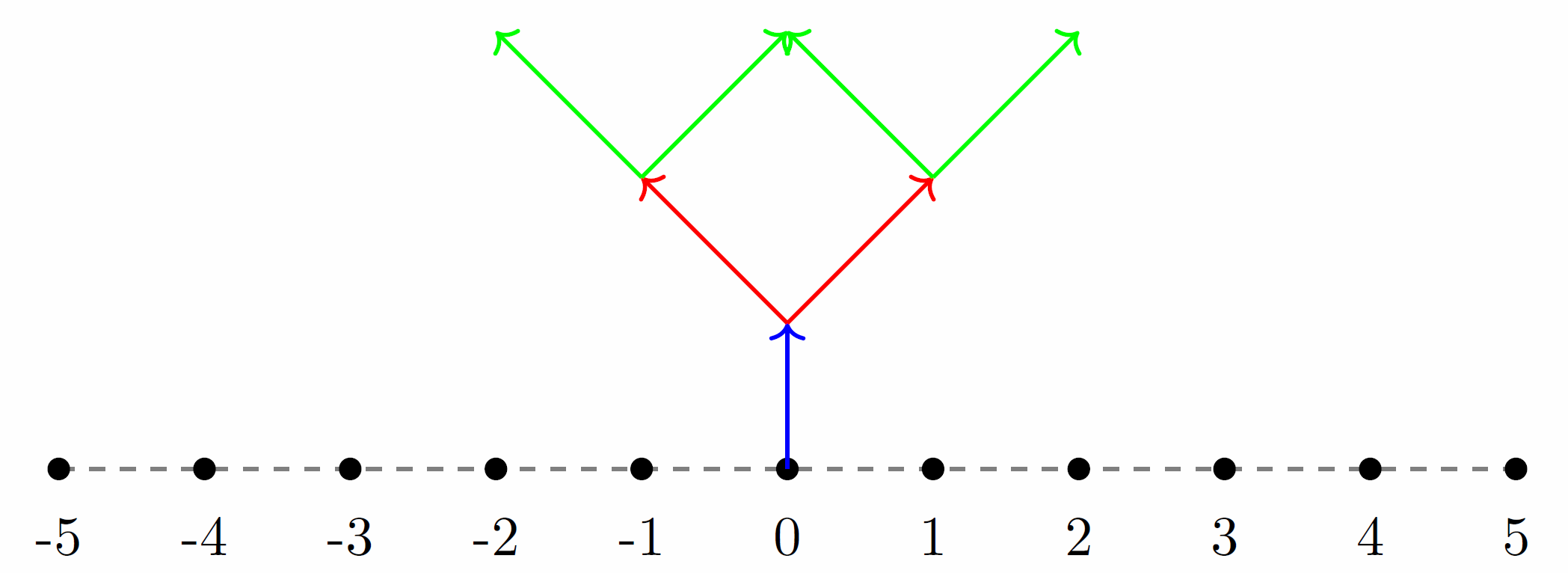}
    \caption{Diagram of a classical random walk in one dimension. The walker starts at position 0 and moves left or right with equal probability at each step.}
    \label{fig:CRW}
\end{figure}

\begin{figure}
    \centering
    \includegraphics[width=0.8\textwidth]{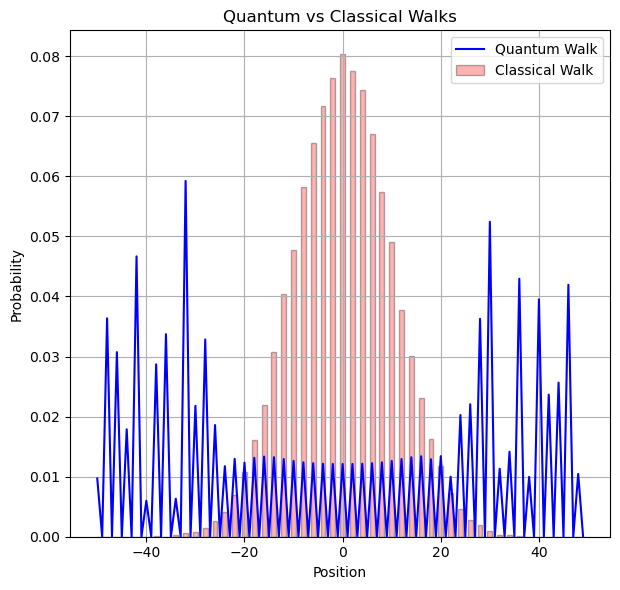}
    \caption{Comparison of probability distributions from classical and quantum random walks. The figure is intended as a qualitative illustration: interference effects in the quantum walk lead to pronounced peaks, in contrast with the smooth, binomial-like spread of the classical walk. Specific simulation details are provided in the accompanying Jupyter notebook\cite{Jupyter}.}
    \label{fig:Distribution}
\end{figure}

Students begin by exploring the classical random walk \cite{pathria2011} which is typically introduced in statistical mechanics courses to illustrate diffusion and stochastic processes. In this model, a particle moves left or right at each time step (Fig.~\ref{fig:CRW}), based on a fair coin toss. After many steps, the particle's position follows a bell-shaped Gaussian distribution, with variance growing linearly with the number of steps. Using Python code provided in the accompanying Jupyter notebook\cite{Jupyter}, students simulate many realizations of the walk and observe the emergence of this characteristic distribution, as seen in Fig.~\ref{fig:Distribution}. They are encouraged to vary the number of steps and trials, and to analyze how the histogram evolves.

% \begin{figure}
%     \centering
%     \begin{tikzpicture}
%     % Draw the line for the 1D lattice
%     \draw[thick, gray, dashed] (-5, 0) -- (5, 0);

%     % Draw the nodes representing positions
%     \foreach \x in {-5,-4,...,5} {
%         \filldraw (\x, 0) circle (2pt);
%     }

%     % Draw labels for positions
%     \foreach \x/\label in {-5/-5, -4/-4, -3/-3, -2/-2, -1/-1, 0/0, 1/1, 2/2, 3/3, 4/4, 5/5} {
%         \node[below] at (\x, -0.2) {\small \label};
%     }

%     % Draw the initial state
%     \draw[->, thick, blue] (0, 0) -- (0, 1);

%     % First level of walk
%     \draw[->, thick, red] (0, 1) -- (-1, 2);
%     \draw[->, thick, red] (0, 1) -- (1, 2);

%     % Second level
%     \draw[->, thick, green] (-1, 2) -- (-2, 3);
%     \draw[->, thick, green] (-1, 2) -- (0, 3);
%     \draw[->, thick, green] (1, 2) -- (0, 3);
%     \draw[->, thick, green] (1, 2) -- (2, 3);
%     \end{tikzpicture}
%     \caption{Diagram of a classical random walk in one dimension. The walker starts at position 0 and moves left or right with equal probability at each step.}
%     \label{fig:CRW}
% \end{figure}

\begin{figure}
    \centering
    \includegraphics[width=0.65\linewidth]{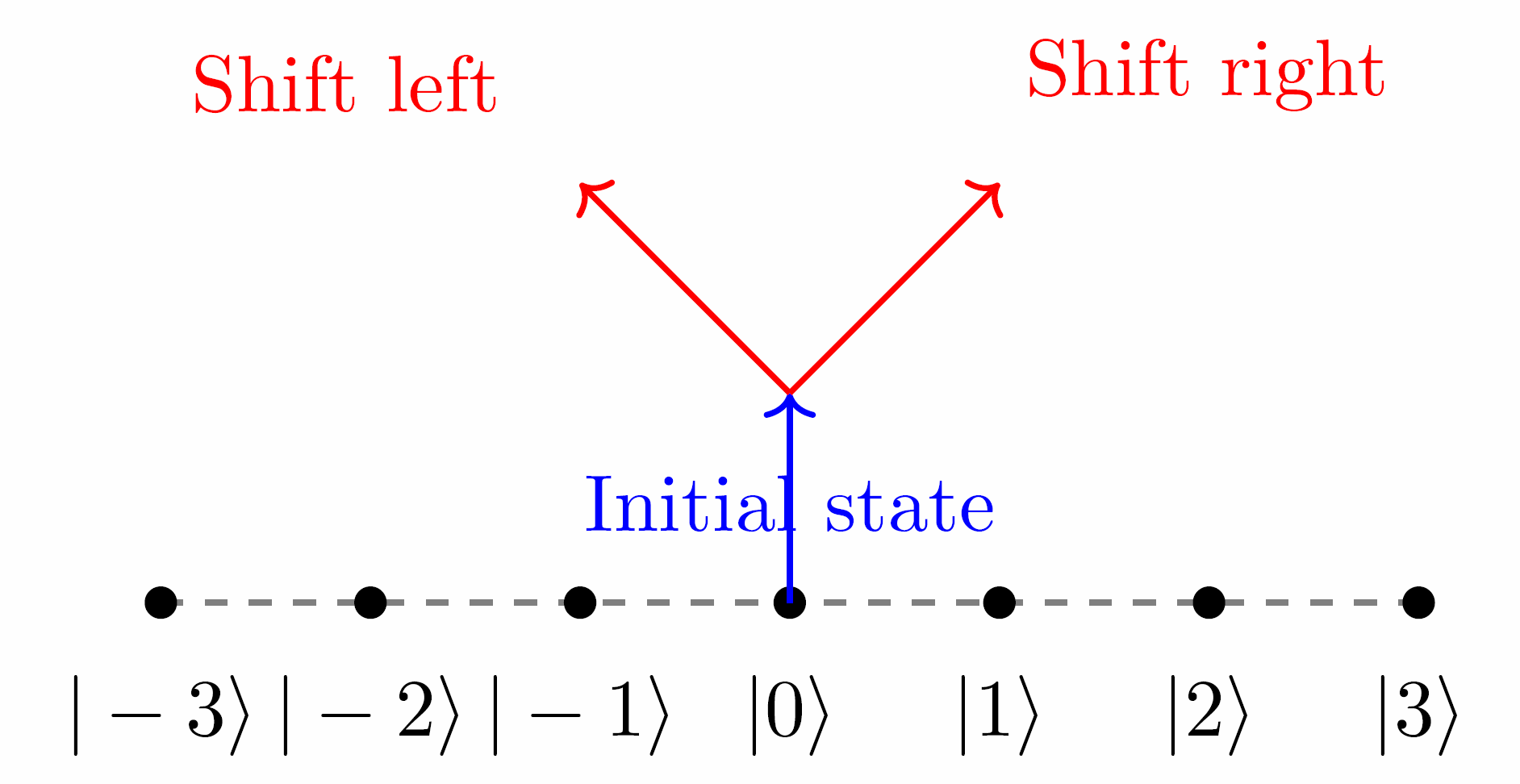}
    \caption{Diagram of a quantum random walk in one dimension. The Hadamard gate creates a superposition state at each step, allowing the walker to explore multiple paths simultaneously. Here, the site index $x$ is represented by a multi-qubit register rather than a single qubit, allowing binary encoding of positions. Thus, the quantum state $\vert x \rangle$ refers to the state of multiple qubits forming a quantum register.}
    \label{fig:QRW}
\end{figure}

The transition to quantum walks \cite{nielsen2010, susskind2019, bouche2021} invites students to rethink randomness. Unlike classical walkers, quantum walkers can occupy a superposition of positions, evolving according to rules dictated by unitary transformations. Here, a Hadamard gate plays the role of a quantum coin, creating a superposition of leftward and rightward moves. After each coin operation, a shift operator moves the particle conditionally: left if the coin qubit is in one state, right if it is in the other, as seen in Fig.~\ref{fig:QRW}. Because each path evolves simultaneously, interference emerges between the various possible outcomes.

% \begin{figure}
%     \centering
%     \begin{tikzpicture}
%         % Draw the line for the 1D lattice
%         \draw[thick, gray, dashed] (-3, 0) -- (3, 0);

%         % Draw nodes representing positions
%         \foreach \x in {-3,-2,-1,0,1,2,3} {
%             \filldraw (\x, 0) circle (2pt);
%         }

%         % Position labels in ket notation
%         \foreach \x/\label in {-3/{\small $|-3\rangle$}, -2/{\small $|-2\rangle$}, -1/{\small $|-1\rangle$}, 0/{\small $|0\rangle$}, 1/{\small $|1\rangle$}, 2/{\small $|2\rangle$}, 3/{\small $|3\rangle$}} {
%             \node[below] at (\x, -0.2) {\label};
%         }

%         % Initial superposition state at origin
%         \node[above, blue] at (0, 0.2) {\small Initial state};
%         \draw[->, thick, blue] (0, 0) -- (0, 1);

%         % First level steps from initial state at origin
%         \node[above right, red] at (-3, 2.2) {\small Shift left};
%         \node[above left, red] at (3, 2.2) {\small Shift right};
%         \draw[->, thick, red] (0, 1) -- (-1, 2);
%         \draw[->, thick, red] (0, 1) -- (1, 2);
%     \end{tikzpicture}
%     \caption{Diagram of a quantum random walk in one dimension. The Hadamard gate creates a superposition state at each step, allowing the walker to explore multiple paths simultaneously. Here, the site index $x$ is represented by a multi-qubit register rather than a single qubit, allowing binary encoding of positions. Thus, the quantum state $\vert x \rangle$ refers to the state of multiple qubits forming a quantum register.}
%     \label{fig:QRW}
% \end{figure}

The resulting distribution of a quantum random walk is dramatically different: instead of the Gaussian profile, students observe a double-peaked structure with a flatter center. These peaks, as seen in Fig.~\ref{fig:Distribution},  are the result of constructive interference, while troughs form where destructive interference occurs. The standard deviation grows linearly rather than with the square root of the number of steps, what we refer to as ballistic spreading. A side-by-side comparison of classical and quantum cases (Table~\ref{tab:comparison}) helps students appreciate these core differences and connects abstract quantum principles with familiar statistical patterns.

Quantum coin operations introduce relative phases between states. In the simplest implementation discussed here, phases are either $0$ or $\pi$. However, other choices of phases are possible and would yield different interference patterns, highlighting the quantum nature of these walks.

\begin{table}
\centering
\begin{tabular}{|l|p{5.1cm}|p{6.5cm}|}
\hline
\textbf{Feature} & \textbf{Classical Random Walk} & \textbf{Quantum Random Walk} \\
\hline
Position Distribution & Gaussian & Non-Gaussian with multiple peaks \\
\hline
Mean Displacement $\langle x_n \rangle$ & 0 & 0 \\
\hline
Variance $\langle x_n^2 \rangle$ & Linear growth ($\sim n$) & Quadratic growth ($\sim n^2$) \\
\hline
Speed of Spread & Diffusive ($\sim \sqrt{n}$) & Ballistic ($\sim n$) \\
\hline
Interference Effects & None & Strong constructive/destructive interference \\
\hline
Applications & Diffusion, Brownian motion, random walks & Quantum algorithms, quantum search, quantum diffusion \\
\hline
\end{tabular}
\caption{Comparison of classical and quantum random walks.}
\label{tab:comparison}
\end{table}

To deepen understanding, we include an optional analytical exercise in ~\ref{appendix:quantumwalk} where students work through the evolution of the quantum state over the first few steps. Starting from an initial superposition state,
\begin{equation}
|\psi_0\rangle = \frac{1}{\sqrt{2}} \left( |\uparrow\rangle \otimes |0\rangle + |\downarrow\rangle \otimes |0\rangle \right),
\end{equation}
they apply the shift operator
\begin{equation}
S = |\uparrow\rangle \langle \uparrow| \otimes \sum_i |i+1\rangle \langle i| + |\downarrow\rangle \langle \downarrow| \otimes \sum_i |i-1\rangle \langle i|,
\end{equation}
and trace how the walker's state evolves.
In practice, each position is encoded using multiple qubits as a quantum register. Each qubit in this register represents a binary digit, enabling the efficient encoding of the walker's position. For example, if the walker spans 8 sites, at least 3 qubits are needed to encode the position states $\ket{0}$ through $\ket{7}$. This binary encoding enables the walker to simultaneously explore multiple paths in superposition. This exercise helps students practice Dirac notation and builds intuition for how quantum interference arises from the superposition of paths.

\begin{figure}
    \centering
    \includegraphics[width=0.8\textwidth]{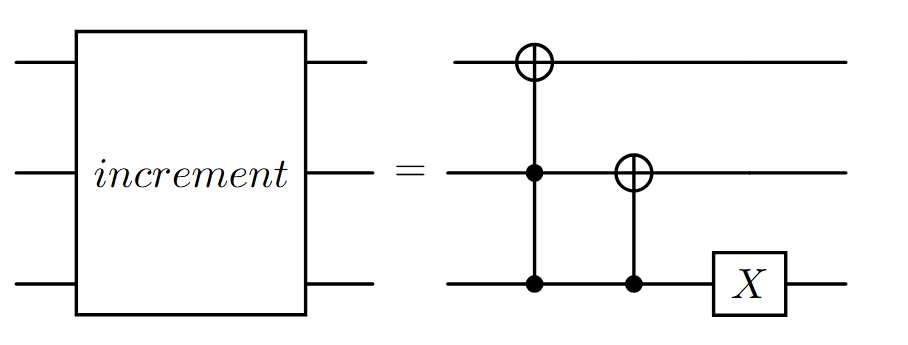}
    \caption{Quantum circuit (3 qubits) to implement movement to the right  in binary form using a CCNOT, a CNOT, and an X gate.}
    \label{fig:right}
\end{figure}

\begin{figure}
    \centering
    \includegraphics[width=0.8\textwidth]{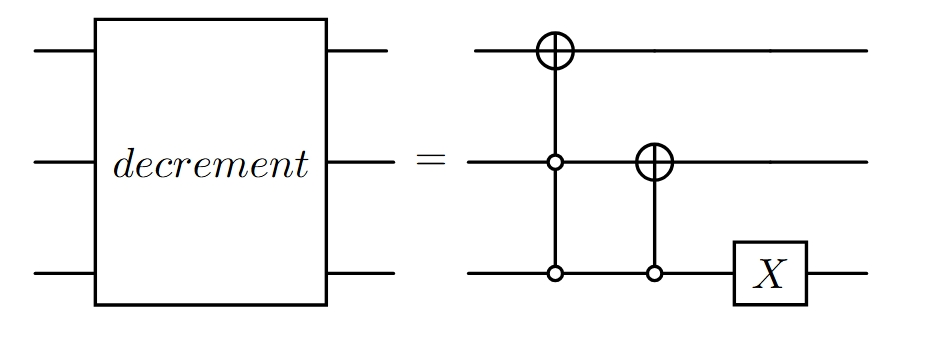}
    \caption{Quantum circuit (3 qubits) to implement movement to the left in binary form using a CCNOT, a CNOT, and an X gate.}
    \label{fig:left}
\end{figure}

\begin{figure}
    \centering
    \includegraphics[width=0.8\textwidth]{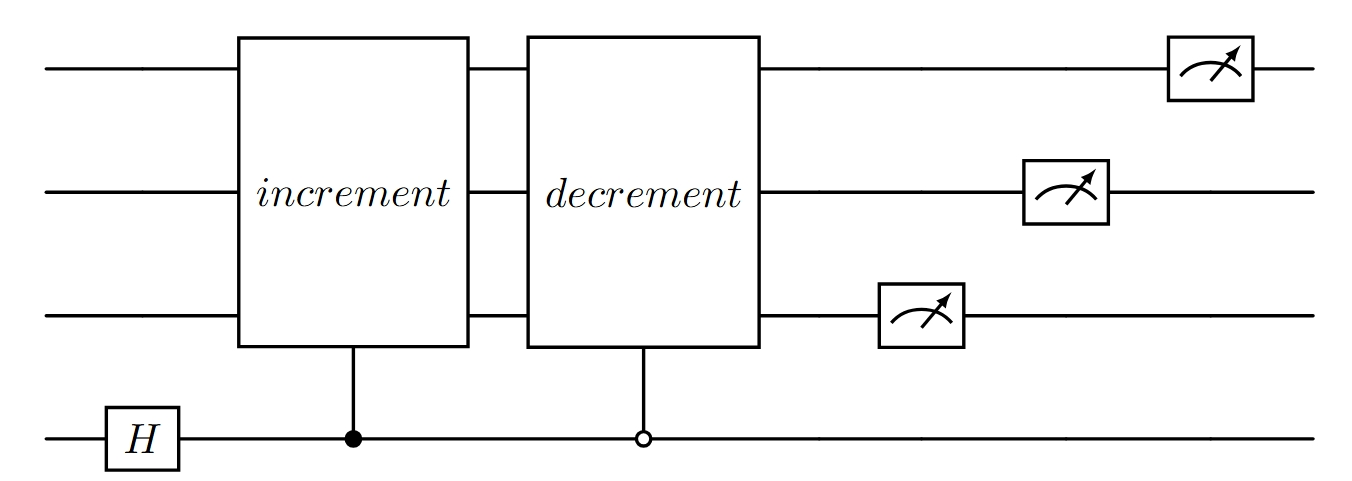}
    \caption{Full quantum walk circuit for unbiased random walk implementing  3 qubits (4 steps on either side).}
    \label{fig:main}
\end{figure}

Having understood the theoretical setup, students implement the quantum random walk using Qiskit in the accompanying Jupyter notebook\cite{Jupyter}. The circuit includes a coin register (a single qubit) and a position register (encoded in binary using multiple qubits). The Hadamard gate initializes the coin in a superposition state. Controlled shift gates then update the position register depending on the coin's value. Figures  ~\ref{fig:right}, ~\ref{fig:left}, and ~\ref{fig:main} show the progression of the quantum circuit for the random walk using 3 qubits.

In the quantum circuits explained in detail in the Jupyter notebook\cite{Jupyter}, the ancilla register is introduced to implement conditional position updates while preserving quantum coherence. Rather than measuring the position qubits directly---an action that would collapse the wavefunction---the ancilla qubits enable controlled operations that shift the walker's position without destroying the quantum state.

After constructing the full circuit, students simulate the walk using Qiskit's Aer backend. They record outcomes after many trials, producing a histogram of final positions. When compared to the classical histogram from earlier in the notebook, the differences are immediately visible, reinforcing both conceptual understanding and the impact of quantum interference.

The probability of finding a quantum walker at a given site is obtained by taking the squared magnitude of the quantum amplitude at that site. This fundamental rule of quantum mechanics is central to understanding the statistical outcomes observed in quantum random walks.

Instructors might encourage students to think critically about the simulations they've explored. For example, what aspects of the quantum probability distribution reveal the presence of interference? Why does the quantum walker, despite its sharply peaked spread, still have an average position of zero? Students can also consider how the classical and quantum walks differ in how their spread scales with time, and begin to speculate about real-world situations, such as physical systems or computational algorithms, where one model might offer an advantage over the other.

It is worth highlighting that the distinctly quantum behavior observed in quantum random walks originates from coherent quantum evolution, allowing interference effects to build up over successive steps---an essential feature lacking a classical analogue. If one instead performed a measurement on the quantum state at each individual step (thus prematurely collapsing the wavefunction), the quantum behavior would disappear. This continuous measurement scenario effectively reduces the quantum random walk to its classical counterpart, underscoring the foundational role of coherence and interference in quantum computation. Clarifying this connection helps demystify quantum computation for students, highlighting precisely how qubit-based algorithms differ fundamentally from classical, bit-based computations.

By the end of this module, students not only understand a key difference between classical and quantum systems, but also gain hands-on experience with quantum circuit construction, simulation, and analysis. These are skills that will serve them well in further explorations of quantum information and statistical physics.

\section{Simulating the Ising Model with Quantum Circuits}
\label{sec:ising}

The Ising model \cite{pathria2011} is a mainstay of statistical physics education, offering students a concrete way to understand how microscopic interactions give rise to macroscopic behavior. In its classical form, it models a system of spins, each either up or down, that interact with their neighbors and respond to thermal fluctuations. Even simple versions of the model capture rich physical phenomena, such as spontaneous magnetization, phase transitions, and critical behavior. Because analytical solutions are rare beyond one dimension, the model is typically taught through simulations \cite{newman1999,binder2010,gould2007}.

Students usually encounter the Ising model via the Metropolis algorithm \cite{metropolis1953}, which provides a computational window into equilibrium statistical mechanics. In our module, we use that classical simulation as a stepping stone toward something more novel: implementing the classical Ising model using quantum circuits. This approach doesn't replace the classical one, it deepens it. Students see how the same physical model can be reformulated using the tools of quantum computing, helping them build fluency in both domains while developing a stronger conceptual grasp of the underlying physics.

The inclusion of the Ising model serves a clear pedagogical purpose: it provides students an explicit and accessible example of how quantum computing can be leveraged to gain insights into classical problems widely taught in statistical physics courses. This reinforces a fundamental educational objective of helping students connect classical concepts with quantum approaches.

\subsection{From Metropolis to Quantum Circuits}
The quantum simulations presented here are based on the algorithms presented in  \cite{cole2004}. We found this paper  to be an approachable  transition from a well-known computational method, the Metropolis algorithm, to a `Metropolis-like' implementation using the tools of quantum computing.

In this activity, we illustrate how quantum circuits can be used to implement a rule-based spin-flip protocol that lowers energy, mimicking a zero-temperature Metropolis update. Full Metropolis dynamics at finite temperature would require encoding thermal probabilities and accepting or rejecting spin flips probabilistically---a nontrivial step not implemented here. Our goal is to introduce students to energy-based conditional logic using quantum gates, while inviting further exploration in more advanced settings.

The Jupyter notebook\cite{Jupyter} that accompanies this module is designed to guide students along this arc, beginning with the classical and ending in the quantum. Students first implement a standard 1D Metropolis simulation in Python. They initialize a chain of spins, define the interaction energy, and write the familiar acceptance rule based on the Boltzmann factor: flips that reduce energy are always accepted; flips that increase energy are accepted with a probability \( e^{-\Delta E / k_B T} \). 

This part of the notebook serves as a purposeful review: it refreshes students' memory of spin dynamics, energy landscapes, and thermal behavior while establishing a clear foundation for the quantum circuit implementation that follows. By plotting average magnetization, energy per spin, or histograms of spin configurations, students visualize how a system evolves toward thermal equilibrium. They can experiment with temperature, system size, or boundary conditions, and begin asking familiar questions: when does long-range order emerge? What role do fluctuations play?

Then the notebook changes gears. It introduces a different way to simulate the same physics, this time using quantum circuits. The transition is framed not as a leap into abstraction, but as a natural progression. Students are told: `We're going to model the Ising system using quantum gates instead of random numbers. Let's see what stays the same, and what has to change.'

\subsection{Encoding Energy and Probability with Quantum Operations}

Each spin in the system is now represented by a qubit, initialized in the state \(|0\rangle\) or \(|1\rangle\) to match classical spin-up or spin-down. 

\begin{figure}
    \centering
    \includegraphics[width=0.8\textwidth]{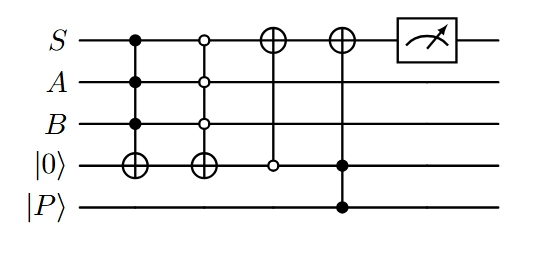}
    \caption{Quantum circuit implemented in Qiskit for the 1D Ising model (no external field), following the method of Cole \textit{et al.}~\cite{cole2004}.}
    \label{fig:ising1D}
\end{figure}

Unlike the classical Metropolis algorithm, which directly implements probabilistic spin flips based on a random number generator and the Boltzmann factor, the quantum circuit version separates the energy evaluation from the acceptance decision. In our notebook, quantum circuits are used to compute the energy contribution of a given spin configuration using controlled gates and measurement. 
In the circuit diagram (Fig.~\ref{fig:ising1D}), gates with black dots represent the traditional CCCX gate (Toffoli-3 gate), where the target qubit is flipped only when all three control qubits are in the \(|1\rangle\) state (in our example, spin-down). Gates with white dots represent the reversed CCCX gate, which flips the target qubit when all three control qubits are in
the \(|0\rangle\) state (spin-up). This circuit is applied to each spin in the system, and the final configurations are recorded. The process is repeated multiple times, similar to the Monte Carlo method. The implementation of this circuit is shown in detail in the Jupyter notebook\cite{Jupyter}.

While Toffoli gates may not be optimal for practical large-scale quantum computations, their inclusion here is deliberate, emphasizing conceptual clarity. For students new to quantum computing, Toffoli gates clearly illustrate logical operations, providing a familiar analogy with classical computing logic, and aligning perfectly with the educational intent of this paper.

Students construct this circuit in stages. The notebook walks them through the initialization of the qubits and the application of controlled gates for spin-spin interactions. Measurement collapses the quantum state and yields a specific spin configuration. By repeating the simulation many times, students build up a histogram of outcomes that reflects the same statistical patterns they observed in the Metropolis simulation.

The value of this activity lies not in outperforming the classical algorithm, but in offering a new way of thinking. Students see that quantum circuits can simulate classical thermal behavior, and begin to understand the cost and complexity of doing so. They also confront the challenges of working with quantum gates and the tradeoffs involved in probabilistic outcomes that emerge only after measurement.

\subsection{Scaling to 2D with simplified layers}

Once students have a feel for the 1D case, the notebook introduces a 2D Ising model. Of course, simulating a full 2D lattice with nearest-neighbor interactions would require deep circuits and many qubits, beyond what most quantum simulators or teaching setups can handle. To address this, we offer a pedagogically motivated simplification: the lattice is split into two 1D chains, one for horizontal interactions and one for vertical. Each chain is simulated as a separate layer in the quantum circuit.

\begin{figure}
    \centering
    \includegraphics[width=0.8\textwidth]{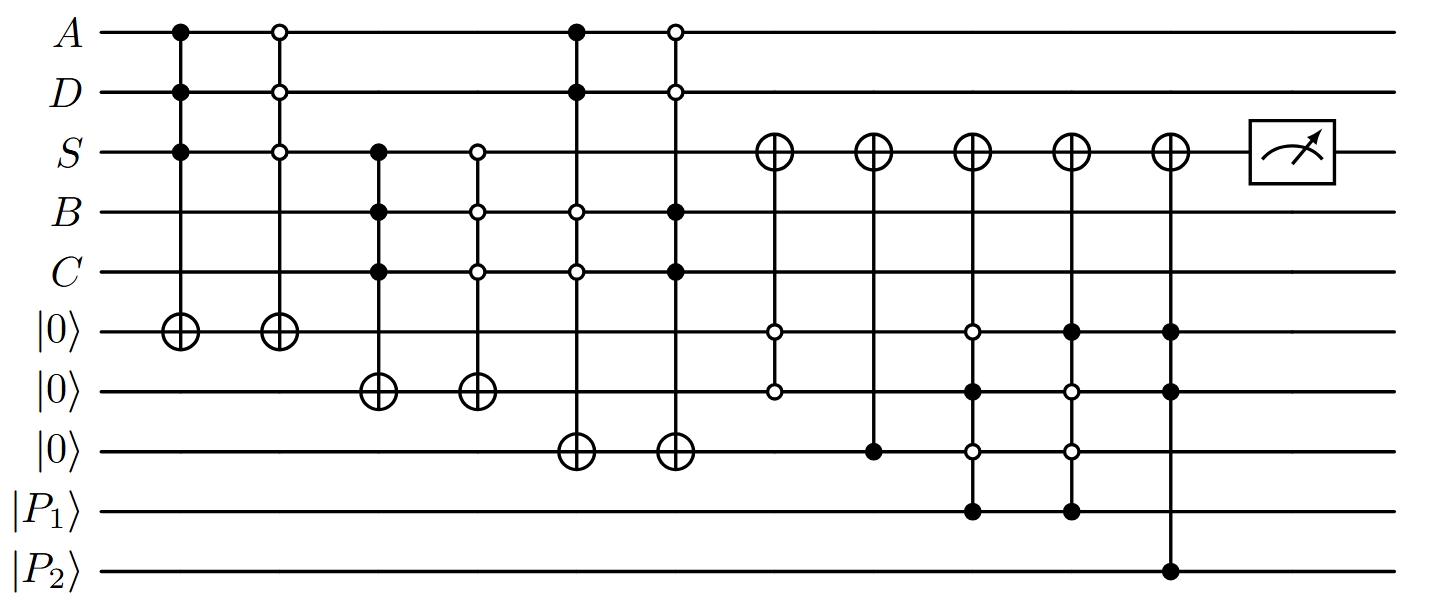}
    \caption{Quantum circuit implemented in Qiskit for the 2D Ising model. Horizontal and vertical interactions are handled in separate circuit layers.}
    \label{fig:ising2D}
\end{figure}

This decomposition keeps the simulation tractable, but still reveals the key features of the 2D system. Students compare the output of the quantum simulation with the classical Metropolis results. They look for patterns in the spin configurations, explore how average magnetization depends on temperature, and reflect on what is preserved and what is lost in the quantum approach. The circuit is presented in Fig.~\ref{fig:ising2D}.

Throughout these exercises, students are prompted to compare the two simulations. How do they model the same physics? Where does the logic differ? What does it mean to simulate thermal probability using quantum amplitudes? This algorithm provides an optimized approach to simulating the two-dimensional classical Ising model using quantum circuits.

For the same problem, the circuit can be further optimized to improve its efficiency. In a two-dimensional classic Ising model, a spin has 4 neighbors, giving a total of \(2^4 = 16\) different combinations, with the central spin's direction fixed. Among all of these combinations, only 5 of them will lead to a positive energy change after a flip, and thus, a probabilistic flip. Compared to the previous algorithm, which calculates the probability of flipping a spin under different neighbors' combinations, it would be more efficient if we consider the probability of NOT flipping a spin if the neighbors are in one of the 5 combinations with positive energy changes. The advantage of this design is that if the spins' configuration will not yield a positive energy change (which is the majority in all the cases), it will not go through this circuit, saving computation time.\

The same principle can be extended to the three-dimensional classical Ising model, as well as to other variants, such as the Ising model with diagonal-neighbor interactions, and lattice structures such as Cayley trees. The details of this circuit optimization are explored step by step in the accompanying Jupyter notebook\cite{Jupyter}. There, students are guided through the full process of constructing the algorithm and assembling the corresponding quantum circuits, gaining both conceptual understanding and practical coding experience along the way.
  
Instructors may pause here to introduce deeper questions: Could this circuit be run on a real quantum computer? What would be the limitations? What other models might be adapted in this way? Using the notebook, students can explore optional challenges, such as adding an external magnetic field, changing interaction signs, or incorporating transverse field terms for quantum annealing \cite{das2008}.

By the end of the module, students have not just learned how to simulate the Ising model using quantum circuits. They have strengthened their understanding of energy, probability, and equilibrium, while gaining hands-on experience in translating those ideas into code. More importantly, they have begun to see that quantum computing is not an isolated subject, but one that intersects meaningfully with the physics they already know.

\section{Conclusions and Future Directions}
\label{sec:conclusion}

This paper has presented two integrated modules for introducing quantum computing concepts in an undergraduate statistical physics course. Each module, one focused on the quantum random walk, the other on the Ising model, uses Qiskit and Jupyter notebooks\cite{Jupyter} to help students explore how classical and quantum systems differ in structure, behavior, and simulation techniques. Rather than teaching quantum computing in isolation, we embed it into a familiar physics context, allowing students to draw meaningful comparisons and deepen their conceptual understanding.

In the quantum walk module, students move from classical diffusion to quantum interference, observing how superposition and unitary evolution shape probability distributions. They construct circuits that mirror coin tosses and shifts, then run them to see how quantum randomness plays out across many trials. Along the way, they learn Dirac notation, circuit diagrams, and Qiskit syntax, all grounded in a topic they already understand from classical physics.

In the Ising model module, students revisit the well-known Metropolis algorithm and then reimplement the same model using quantum circuits. This translation requires them to think carefully about how energy, probability, and thermal behavior can be encoded in  quantum operations. Students learn that while quantum circuits can emulate classical systems, they do so using fundamentally different tools such as gates, amplitudes, and measurement-based probabilities. Through side-by-side simulations and guided exploration in the notebook, students gain both practical skills and conceptual insights.

These modules are designed to be adaptable. Instructors may include them in a statistical mechanics course, a quantum computing elective, or a modern physics sequence. They can be scaled up or down depending on the available time and student preparation. More advanced students might explore quantum annealing via the transverse-field Ising model, or investigate how Grover's algorithm \cite{grover1996} relates to quantum walks. Others might experiment with running the circuits on real quantum hardware through IBM's cloud interface and reflect on noise and error.

Looking ahead, the boundary between classical and quantum modeling will only continue to blur. Our goal is to help students cross that boundary with confidence by showing that the tools of quantum information are not separate from physics, but a natural extension of it. By engaging with quantum circuits through familiar examples like the random walk and the Ising model, students not only develop coding skills and quantum intuition, but also rediscover core ideas in statistical physics from a fresh and exciting perspective.

\appendix
\section{Step-by-step evolution in the quantum random walk}
\label{appendix:quantumwalk}

Let's walk through some basic steps of the quantum random walk process. Let's assume that the walker starts at the origin, \( |0\rangle \), with the coin state initially set to \( |\uparrow\rangle \). This coin state corresponds to a definite direction, such as `right'. First, the  Hadamard gate (\( H \)) is applied to the coin state to create a superposition. The transformation is as follows:
\begin{equation}
H|\uparrow\rangle = \frac{1}{\sqrt{2}}(|\uparrow\rangle + |\downarrow\rangle)
\end{equation}
After applying the Hadamard gate, the overall state of the system becomes:
\begin{equation}
|\psi\rangle = \frac{1}{\sqrt{2}}( |\uparrow\rangle\otimes |0\rangle+ |\downarrow\rangle \otimes |0\rangle)
\end{equation}
This superposition indicates that the walker is in both a state of moving left (\(|\downarrow\rangle \)) and right (\(|\uparrow\rangle \)) simultaneously.

Next we discuss the shift operator. In a quantum random walk, the shift operator \( S \) is used to move the `walker' (or particle) based on the state of the `coin' qubit. The shift operator \( S \) is defined as:
\begin{equation}
S = | \uparrow \rangle \langle \uparrow | \otimes \sum_{i} | i + 1 \rangle \langle i | + | \downarrow \rangle \langle \downarrow | \otimes \sum_{i} | i - 1 \rangle \langle i |.
\end{equation}

This operator consists of two main parts, corresponding to the two possible states of the coin qubit, \( | \uparrow \rangle \) and \( | \downarrow \rangle \). The action of  \( S \) can be understood as follows.

The first term in \( S \) is $| \uparrow \rangle \langle \uparrow | \otimes \sum_{i} | i + 1 \rangle \langle i |$. The expression \( | \uparrow \rangle \langle \uparrow | \) is a projection operator on the coin qubit. It acts as an identity on the \( | \uparrow \rangle \) state and zero on the \( | \downarrow \rangle \) state. The sum \( \sum_{i} | i + 1 \rangle \langle i | \) is an operator that shifts the position state by +1. This operator takes any position state \( | i \rangle \) and maps it to \( | i + 1 \rangle \), effectively moving the walker one step to the right. So, if the coin is in the state \( | \uparrow \rangle \), the operator \( S \) will shift the position \( | i \rangle \) to \( | i + 1 \rangle \). The same reasoning applies to the second term  of the operator. If the coin is in the state \( | \downarrow \rangle \), the operator \( S \) will shift the position \( | i \rangle \) to \( | i - 1 \rangle \).

If we start with a superposition of coin and position states such as:
\begin{equation}
|\psi\rangle = \frac{1}{\sqrt{2}} \left( |\uparrow\rangle \otimes |0\rangle + |\downarrow\rangle \otimes |0\rangle \right),
\end{equation}
then applying \( S \) will operate on each term in this superposition according to the coin state, as described below.

 For \( |\uparrow\rangle \otimes |0\rangle \), the \( | \uparrow \rangle \langle \uparrow | \) projection will act on \( |\uparrow\rangle \), allowing the shift \( \sum_i |i + 1 \rangle \langle i | \) to take effect. This shifts \( |0\rangle \) to \( |1\rangle \), so we get \( |\uparrow\rangle \otimes |1\rangle \).
 For \( |\downarrow\rangle \otimes |0\rangle \), the \( | \downarrow \rangle \langle \downarrow | \) projection will act on \( |\downarrow\rangle \), allowing the shift \( \sum_i |i - 1 \rangle \langle i | \) to take effect. This shifts \( |0\rangle \) to \( |-1\rangle \), so we get \( |\downarrow\rangle \otimes |-1\rangle \).

Combining these results, we get:
\begin{equation}
S |\psi\rangle = \frac{1}{\sqrt{2}} \left( |\uparrow\rangle \otimes |1\rangle + |\downarrow\rangle \otimes |-1\rangle \right).
\end{equation}

 We can combine these two operators (the coin and the shift operators) into one operator $U$ whose action can be represented in a simplified way \cite {bouche2021}:

\begin{equation}
U|\uparrow n\rangle = |\uparrow n+1\rangle + |\downarrow n-1\rangle
\end{equation}
\begin{equation}
U|\downarrow n\rangle = |\uparrow n+1\rangle - |\downarrow n-1\rangle
\end{equation}
where we use the condensed notation $|\uparrow n\rangle =|\uparrow \rangle  \otimes  | n\rangle$.

In order to better understand the quantum interference effects, we propose the following exercise, similar to the example presented in \cite{bouche2021}. This calculation offers students valuable practice with the Dirac notation and application of operators. To illustrate the process, we will calculate the first four steps of a quantum random walk, starting from the initial state \(|\uparrow 0\rangle \). Instances of destructive and constructive interference are  in \textcolor{red}{red} and \textcolor{green}{green}.

\begin{itemize}
    \item Initial state: $|\uparrow 0\rangle$
    
    \item Step 1: $\frac{1}{\sqrt{2}} (|\uparrow 1\rangle + |\downarrow-1\rangle)$
    
    \item Step 2: $\frac{1}{2} (|\uparrow 2\rangle + |\downarrow 0\rangle+|\uparrow 0\rangle-|\downarrow-2\rangle)$
    
    \item Step 3: $\frac{1}{2\sqrt{2}} (|\uparrow 3\rangle + |\downarrow 1\rangle + \textcolor{green}{|\uparrow 1\rangle} - \textcolor{red}{|\downarrow -1\rangle} + \textcolor{green}{|\uparrow 1\rangle} + \textcolor{red}{|\downarrow -1\rangle} - |\uparrow -1\rangle + |\downarrow -3\rangle) = \frac{1}{2\sqrt{2}}(|\uparrow 3\rangle + |\downarrow 1\rangle + 2|\uparrow 1\rangle - |\uparrow -1\rangle + |\downarrow -3\rangle)$
    
    \item Step 4: $\frac{1}{4} (|\uparrow 4\rangle + |\downarrow 2\rangle + \textcolor{green}{|\uparrow 2\rangle} -\textcolor{red}{|\downarrow 0\rangle} + \textcolor{green}{|\uparrow 2\rangle} + \textcolor{red}{|\downarrow 0\rangle} + \textcolor{green}{|\uparrow 2\rangle} + |\downarrow 0\rangle - |\uparrow 0\rangle - |\downarrow -2\rangle + |\uparrow -2\rangle - |\downarrow -4\rangle) = \frac{1}{4} (|\uparrow 4\rangle + |\downarrow 2\rangle + 3|\uparrow 2\rangle + |\downarrow 0\rangle - |\uparrow 0\rangle - |\downarrow -2\rangle + |\uparrow -2\rangle - |\downarrow -4\rangle)$
\end{itemize}

\end{document}